\documentclass[aps,prb,twocolumn,superscriptaddress]{revtex4-1}
\usepackage{graphicx}
\usepackage{dcolumn}% Align table columns on decimal point
\usepackage{bm}% bold math
\usepackage{subfigure}
\usepackage{amssymb}
\usepackage{notoccite}

\newcommand{\pcmo}{Pr$_{0.5}$Ca$_{0.5}$MnO$_{3}$}
\newcommand{\ramox}{R$_{1-x}$A$_{x}$MnO$_{3}$}

\usepackage[usenames,dvipsnames]{color}

\def\ignorecitefornumbering#1{%
     \begingroup
         \@fileswfalse
         #1%                     % do \cite comand
    \endgroup
}

\begin{document}

%Title of paper
\title{Spin excitations used to probe the nature of the exchange coupling in the magnetically ordered ground state of \pcmo}

\author{R. A. Ewings}
\email[]{russell.ewings@stfc.ac.uk}
\affiliation{ISIS Facility,
        STFC Rutherford Appleton Laboratory,
        Harwell Oxford, Didcot OX11 0QX, United Kingdom}

\author{T. G. Perring}
\affiliation{ISIS Facility,
        STFC Rutherford Appleton Laboratory,
        Harwell Oxford, Didcot OX11 0QX, United Kingdom}
\affiliation{London Centre for Nanotechnology, 17-19 Gordon Street, London WC1H 0AH, United Kingdom}

\author{O. Sikora}
\affiliation{H. H. Wills Physics Laboratory,
        University of Bristol,
        Bristol BS8 1TL, United Kingdom}
\altaffiliation[Current address: ]{Marian Smoluchowski Institute of Physics, Jagiellonian University, Prof. {\L}ojasiewicza 11, PL-30348 Krak\'{o}w, Poland}

\author{D. L. Abernathy}
\affiliation{Quantum Condensed Matter Division, Oak Ridge National Laboratory,
        Oak Ridge,
        Tennessee 37831, USA}

\author{Y. Tomioka}
\affiliation{Electronics and Photonics Research Institute, National Institute of Advanced Industrial Science and Technology (AIST), Tsukuba Central 4, 1-1-1 Higashi Tsukuba 305-8562, Japan}

%\author{J. P. Hill}
%\affiliation{Condensed Matter Physics and Materials Science Department, Brookhaven National Laboratory, Upton, New York 11973, USA}

\author{Y. Tokura}
\affiliation{Department of Applied Physics, University of Tokyo, Bunkyo-ku, Tokyo 113-8656, Japan}
\affiliation{RIKEN Center for Emergent Matter Science (CEMS), Wako 351-0198, Japan}

\date{\today}

\begin{abstract}
% insert abstract here
We have used time-of-flight inelastic neutron scattering to measure the spin wave spectrum of the canonical half-doped manganite \pcmo\, in its magnetic and orbitally ordered phase. The data, which cover multiple Brillouin zones and the entire energy range of the excitations, are compared with several different models that are all consistent with the CE-type magnetic order, but arise through different exchange coupling schemes. The Goodenough model, i.e. an ordered state comprising strong nearest neighbor ferromagnetic interactions along zig-zag chains with antiferromagnetic inter-chain coupling, provides the best description of the data, provided that further neighbor interactions along the chains are included. We are able to rule out a coupling scheme involving formation of strongly bound ferromagnetic dimers, i.e. Zener polarons, on the basis of gross features of the observed spin wave spectrum. A model with weaker dimerization reproduces the observed dispersion but can be ruled out on the basis of discrepancies between the calculated and observed structure factors at certain positions in reciprocal space. Adding further neighbor interactions results in almost no dimerization, i.e. recovery of the Goodenough model. These results are consistent with theoretical analysis of the degenerate double exchange model for half-doping, and provide a recipe for how to interpret future measurements away from half-doping, where degenerate double exchange models predict more complex ground states.
\end{abstract}

%Provisional choice of PACS numbers
%\pacs{75.30.Ds,75.47.Lx,78.70.Nx,75.25.-j}

\maketitle

\section{Introduction}{\label{sec:intro}}

A characteristic of many materials exhibiting strong electronic correlations is a delicate balance between multiple competing phases which can often be tuned with relatively modest changes in external parameters such as temperature, magnetic field, pressure or chemical doping \cite{Dagotto-2005}. Preeminent amongst strongly correlated electron systems are transition metal oxides, which display phenomena as diverse as superconductivity \cite{Orenstein-2000}, orbital order \cite{Tokura-Nagaosa-2000}, and complex magnetic order \cite{radaelli1997}. Most families of transition metal oxides have been studied extensively, and manganites are no exception.

Interest in the manganites in recent times has broadly centered on two phenomena -- colossal magnetoresistance (CMR) \cite{tokura2006} and multiferroicity \cite{Kimura-2003}. CMR manganites are in many ways typical of strongly correlated electron systems, in that they exist at the cusp of several competing phases \cite{Moreo-1999}. For the particular case of (nearly) cubic \ramox (where R is a rare-earth or La, and A is Ca, Sr, Ba, or Pb) and its layered analogs (R,A)$_{2}$MnO$_{4}$ and (R,A)$_{3}$Mn$_{2}$O$_{7}$ such phases include charge-ordered, antiferromagnetic, ferromagnetic metal, and ferromagnetic insulator. In these perovskite manganites each Mn ion has three electrons in the $t_{2g}$ $d$-orbitals, and a fraction $(1-x)$ have one in the higher energy $e_{g}$ orbitals, whose spins are aligned parallel to the local $t_{2g}$ moment by strong intrasite exchange. The essential physics is a competition between delocalization of the $e_{g}$ electrons, favoring a homogeneous, fully spin polarized, ferromagnetic metal via double exchange, and localization due to lattice distortions trapping the $e_{g}$ electrons to form polarons \cite{Millis98}. For $x\sim0.3$ the materials are typically ferromagnetic metals, with a sea of polarons in the insulating paramagnetic phase. The most marked CMR effects often occur for those materials in which the polarons order in the half doped state, $x\sim0.5$, to give a charge and orbitally ordered insulating state for $T < T_{\rm{CO}}$, the charge-ordering temperature. In the case of \pcmo (PCMO) \,\,$T_{\rm{CO}}=260$\,K and the transition from a paramagnetic to antiferromagnetic phase occurs at $T_{\rm{N}}=180$\,K \cite{Mori-1999}.

\begin{figure*}[ht]\includegraphics*[scale=0.65,angle=0]{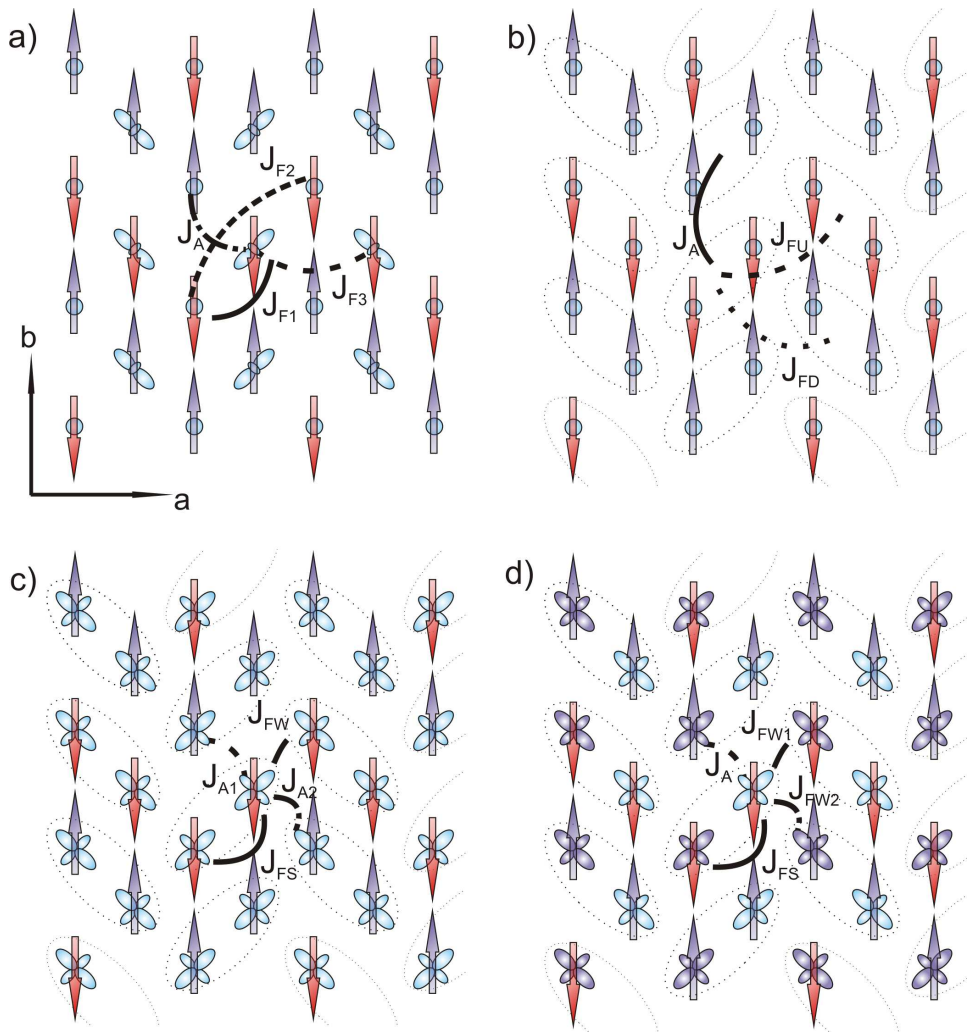}
\centering \caption{(Color online) (a) Illustration of the Goodenough model (see text).  The spins in different one-dimensional chains (zig-zags) are indicated by either red (dark) or purple (light) arrows. The orbitals on the Mn3+ sites are $d_{3x^{2}-r^{2}}$ or $d_{3y^{2}-r^{2}}$ in character, where $x$ and $y$ point along the two perpendicular Mn-Mn directions along the zig-zags. In the model as originally proposed by Goodenough\cite{goodenough1955} there is a ferromagnetic interaction between nearest neighbors along a zig-zag, J$_{F1}$, and antiferromagnetic coupling between nearest neighbors in adjacent zig-zags, J$_{A}$. To explain our data it is necessary to allow for second neighbor interactions, J$_{F2}$ and J$_{F3}$, within zig-zags (see main text). Nearest neighbor antiferromagnetic coupling between planes, J$_{c}$, is not shown. (b) The ZP model, indicating exchange interactions between rigidly coupled $S=7/2$ units in the $ab$-plane. The ZP units are indicated by dashed ovals. Parallel units have AFM coupling, J$_{A}$, and perpendicular units have ferromagnetic couplings, either J$_{FU}$ or J$_{FD}$.  (c) The dimer model, as proposed by Johnstone {\it et al} \cite{Johnstone-2012}, in which spins are not rigidly coupled as in the ZP case, but rather have a stronger ferromagnetic intra-dimer coupling, J$_{FS}$, and weaker inter-dimer couplings, FM J$_{FW}$, and AFM J$_{A1}$ and J$_{A2}$. The orbitals are shown to have a mixed character of a superposition of $d_{y^{2}-z^{2}}$ ($d_{x^{2}-z^{2}}$) and $d_{3x^{2}-r^{2}}$ ($d_{3y^{2}-r^{2}}$), following the conventions used in model calculations for extensions to the DDEX model \cite{Barone2011,Yamauchi-JPSJ-2014} discussed in the text. As originally proposed the model was purely phenomenological without reference to an electronic model, and the orbitals should be considered purely schematic in this panel. (d) An alternative more general dimer model (see text) in which we distinguish between the Mn sites with different proportions of $d_{y^{2}-z^{2}}$ ($d_{x^{2}-z^{2}}$) and $d_{3x^{2}-r^{2}}$ ($d_{3y^{2}-r^{2}}$) character, indicated here by purple and light blue coloring respectively. The orbital character together with Goodenough-Kanamori rules lead to FM J$_{FW2}$, and AFM J$_{A}$, different to the model shown in (c). } \label{fig:scheme_def}
\end{figure*}

Despite its relevance to the defining properties of the CMR manganites, the nature of the charge-ordered state is not yet well-understood. At half-doping, the picture widely assumed is of CE-type charge and magnetic ordering, illustrated in Fig. \ref{fig:scheme_def}(a), first proposed by Goodenough\cite{goodenough1955}. In this original picture, half the Mn ions are Mn$^{3+}$ (hence spin 2) and half are Mn$^{4+}$ (spin 3/2) in a plane of the pseudocubic lattice; the coupling is ferromagnetic (FM) between ions  along the zig-zag chains, with antiferromagnetic (AFM) inter-chain coupling. Along the zig-zag chains the nearest neighbor Mn sites are occupied by alternating Mn$^{3+}$ and Mn$^{4+}$ ions, with the lobes of the d$_{3z^{2}-r^{2}}$-type orbitals on the Mn$^{3+}$ sites pointing at the Mn$^{4+}$ ions on the zig-zags. These planes are then stacked antiferromagnetically. There is considerable experimental evidence in favor of the Goodenough model. Numerous diffraction and x-ray resonant scattering results point to two inequivalent Mn sites in various half-doped manganites \cite{larochelle-2005,Herrero-Martin2004,garcia2001,radaelli1997,goff2004}, and the CE-type magnetic structure is universally accepted \cite{Sternlieb1996,radaelli1997,jirak}. The charge modulation is markedly less than expected in this simple picture in various half-doped manganites, however, including La$_{0.5}$Sr$_{1.5}$MnO$_{4}$, Nd$_{0.5}$Sr$_{0.5}$MnO$_{3}$ and \pcmo. Bond valence sums applied to the Mn ions typically reveal Mn valences of approximately $3.4+$ and $3.6+$ rather than the expected $3+$ and $4+$ at the two sites \cite{goff2004,radaelli1997}, and X-ray resonant scattering at the Mn K-edge also suggests that the valences differ by the same amount \cite{Herrero-Martin2004,garcia2001}.

The formation of the charge-ordered and CE-type magnetic structure at half-doping can be understood theoretically in terms of the degenerate double exchange (DDEX) model \cite{vdBrink-1999,efremov2004}. The DDEX model is a generalization of double exchange that allows for different occupancies of the two-fold degenerate $e_{g}$ orbitals on a lattice of Mn sites, finite Hund's coupling between the $e_{g}$ and $t_{2g}$ orbitals, and superexchange between the core moments of the Mn sites. Frustration between double exchange, which favors delocalization and ferromagnetic alignment of the Mn core moments, and the antiferromagnetic superexchange results in a cooperative orbital and charge ordering that favors ferromagnetism along zig-zag chains and antiferromagnetic stacking of the chains.

At half-doping and for suitable choice of (renormalized) superexchange and electron hopping energies, the DDEX model predicts site-centered charge ordering, with charge disproportionation of $\sim0.1e$ which is similar to the experimental values. Below half-doping there is a range of hole doping, $x$, over which ferromagnetic dimer formation takes place in the ground state. The limiting case of bond-centered charge ordering, in which the Mn ions in a dimer are equivalent, occurs for $x\sim0.4$ . This limiting case of a dimer resembles the Zener polaron \cite{Zener-1951,ZhouGoodenough2000} (ZP), in which adjacent Mn ions pair up via delocalization of the lone $e_{g}$ electron to produce a single strongly bound unit with spin 7/2. The Zener polarons form a herringbone pattern, as shown in Fig. \ref{fig:scheme_def}(b),  with AFM  coupling between parallel units, and FM coupling between perpendicular units.

The idea of the ZP came to prominence following a key single crystal neutron diffraction experiment \cite{daoud-aladine2002} on Pr$_{0.6}$Ca$_{0.4}$MnO$_{3}$ performed at a temperature below $T_{\rm{CO}}$ in which the orientation of the elongated MnO$_{6}$ octahedra and the two similar (although not identical) Mn sites in the structural refinement suggest the formation of Mn--Mn pairs. The concept of Zener polarons also received theoretical support from {\it ab initio} calculations performed at the time in half-doped manganites \cite{Zheng2003,patterson05,ferrari2003}. However, the existence of Zener polarons has proved controversial. Notwithstanding the evidence in support of the Goodenough model for half-doped manganites cited above, X-ray resonant scattering results point to two inequivalent Mn sites in Pr$_{0.6}$Ca$_{0.4}$MnO$_{3}$ itself \cite{grenier2004}, which in the full analysis is shown not to be consistent with the form of ZP defined in ref. \onlinecite{daoud-aladine2002}. Indeed, of the studies cited in favor of the Goodenough model \cite{larochelle-2005,Herrero-Martin2004,garcia2001,goff2004}, many explicitly rule out the existence of Zener polarons. On the other hand, reports of high resolution transmission electron microscopy and electron diffraction experiments claim to confirm the existence of ZP-type ordering in Pr$_{1-x}$Ca$_{x}$MnO$_{3}$, $x=0.3-0.5$, with phase coexistence of the ZP and the conventional charge-ordered structure \cite{Jooss2006,Wu2007}.

%Use \ignorecitefornumbering in Latex+Bibtex+Latex+Latex command sequence. Then replace with \cite and re-run with Latex+Latex, to get correct ordering of references

The existence of two inequivalent sites discussed above as evidence against ZP formation in itself does not preclude the existence of dimers made from adjacent Mn ions. There are a continuum of possibilities for hole doping between the ZP (bond centered) and the Goodenough (site centered) limits, in which the two Mn sites in the ferromagnetic dimers that the DDEX model predicts are inequivalent, with different occupancy of the two $e_{g}$ orbitals at the two positions. The dimers break inversion symmetry and this leads to the possibility of ferroelectricity in manganites; indeed there have been reports \cite{Lopes2008,Shukla2014} of spontaneous electric polarization in Pr$_{1-x}$Ca$_{x}$MnO$_{3}$ for a range of $x$, including $x=0.5$. Motivated by the possibility of multiferroicity, more sophisticated theories that extend the DDEX model to incorporate on-site Coulomb repulsion together with MnO$_{6}$ tilting, and / or Jahn-Teller distortions have been developed, which have been analysed in terms of model calculations or a combination of these with DFT for half-doped manganites \cite{Giovannetti2009,Barone2011}. These approaches reveal instability towards some degree of Mn-Mn dimerization for a range of parameters in the model, as do full {\it ab initio} calculations \cite{Colizzi2010,Yamauchi2013} for \pcmo.

Because the origin of the ferromagnetic interaction is double exchange, the characteristic strength of the magnetic exchange can reasonably be expected to be that typical of ferromagnetic manganites \cite{Hwang-98,Ye-2006,Perring-2001,hirota1}, which in turn is consistent with estimates made from realistic values for the band widths \cite{Perring-2001,Shannon-Chatterji-2002}. The values of the magnetic exchange constants within and between dimers will in general be different because of the inequivalent $e_{g}$ orbital occupancies on the two Mn sites. In the extensions to the DDEX model that incorporate interaction with the lattice, Jahn-Teller distortions (which couple with the $e_{g}$ orbital occupancies) and alternating modulation of Mn-O-Mn bond angles also lead to different intra and inter dimer exchange.

The motivation for the current work is two-fold. First is the question of whether or not the Mn ions are paired in a dimer-like fashion in \pcmo. On the basis of previous inelastic neutron scattering measurements of other half-doped manganites \cite{senff1,Ulbrich2011,Johnstone-2012} we might expect the Goodenough model to provide a better description than models involving dimerization. However, since Pr$_{1-x}$Ca$_{x}$MnO$_{3}$ is the material in which the possible existence of Zener polarons came to prominence, we believe that it is important to confirm whether or not this is the case. This then leads to our second motivation, which is to present an exhaustive account of how to distinguish between the various scenarios that arise within DDEX models from an investigation of the spin waves. The measurement of the spin wave excitation spectra using neutron scattering directly yields information about the strength of the magnetic interactions in a material. In addition to the strength of interactions defining the energy of features in the dispersion relations such as overall bandwidth and sizes of gaps, the directions and distances of the interactions define the periodicity in wavevector of those features. These periodicities can be used to distinguish between the Goodenough model, the case of strongly bound dimers (which we refer to as the 'ZP model') and weakly dimerized models ('dimer models'), all of which exhibit the same magnetic structure but arise due to different sets of exchange interactions.

%The question of whether or not the Mn ions form magnetic dimers in half-doped manganites is the motivation for the current work. The measurement of the spin wave excitation spectra using neutron scattering directly yields information about the strength of the magnetic interactions in a material. In addition to the strength of interactions defining the energy of features in the dispersion relations such as overall bandwidth and size of gaps, the directions and distances of the interactions define the periodicity in wavevector of those features.

\begin{figure}[h]
\includegraphics*[scale=0.38,angle=0]{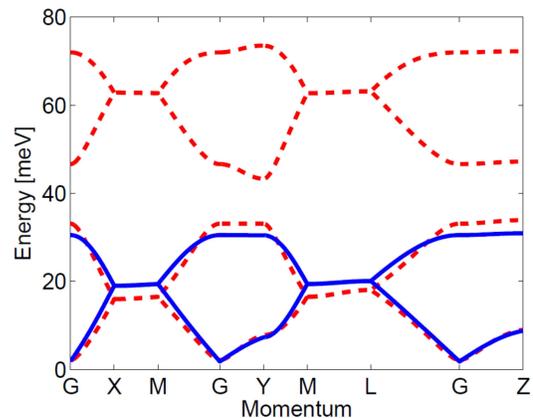}
\centering \caption{ Calculated dispersion relations for the Goodenough model (dashed red line) and ZP model (solid blue line). The parameters for the Goodenough model are those obtained as the global best fit in our analysis (see main text). The parameters for the ZP model are J$_{FU}=1.6$\,meV, J$_{FD}=-0.11$\,meV, J$_{A}=0.11$\,meV and the inter-planar coupling is 1.05\,meV. These parameters were chosen to show that there exists a set of parameters for the ZP model such that at low energies it is practically indistinguishable from the Goodenough model, but that the dispersion at higher energies serves as an effective discriminant. The x-axis labels denote high-symmetry positions in the first Brillouin zone - $G=(0,0,0)$, $X=(\frac{1}{4},0,0)$, $Y=(0,\frac{1}{4},0)$, $Z=(0,0,\frac{1}{2})$, $M=(\frac{1}{4},\frac{1}{4},0)$, and $L=(\frac{1}{4},\frac{1}{4},\frac{1}{2})$. Note that we have chosen to index reciprocal space according to space group $Pbnm$, described in the main text.} \label{fig:olga_dispersion}
\end{figure}

As discussed briefly above, in two earlier studies of the spin waves in CE-type half-doped manganites \cite{senff1,Ulbrich2011}, single-layered La$_{1/2}$Sr$_{3/2}$MnO$_{4}$ and pseudo-cubic Nd$_{0.5}$Sr$_{0.5}$MnO$_{3}$, the lower part of the spin wave dispersion was measured and found to be well-explained by the Goodenough model including second neighbor exchange J$_{F2}$ along the FM zig-zags (see Fig. \ref{fig:scheme_def}(a)). The limiting ZP model of strongly bound dimers, that is, when the intra-dimer exchange is overwhelmingly stronger than any other interactions in the system, was dismissed. The authors argued that of the nearest-neighbor inter-dimer exchange interactions in the ZP model, only that along the AF stacking direction results in significant dispersion because the other two (labeled by J$_{FU}$ within a chain and J$_{FD}$ between chains in Fig. \ref{fig:scheme_def}(b)) are frustrated, resulting in weak or no dispersion perpendicular to the AF stacking direction. This may be the case for the non-collinear structure proposed in ref. \onlinecite{efremov2004} for bond-centered charge ordering on the basis of symmetry, in which the pairs of spins in adjacent ferromagnetic dimers along the zig-zag structure are orthogonal. However, for the case of the known CE-type magnetic structure
\cite{Sternlieb1996,radaelli1997,jirak}, there is no such {\it a-priori} reason to be confident that J$_{FU}$ and J$_{FD}$ are similar. The importance of measuring the full set of dispersion relations is illustrated in Fig. \ref{fig:olga_dispersion}. Considering still the limiting case of strongly bound dimers, Fig. \ref{fig:olga_dispersion} shows theoretical spin wave dispersion relations for the Goodenough model (dashed line) and ZP model (solid line) \footnote{Note that from this point onwards what we refer to as the ZP model is when the intra-dimer exchange is markedly stronger than all other exchange interactions.} At low energies the dispersion relations near the zone center are qualitatively very similar. The main difference, however, is the absence of the higher branches for the ZP model, because the binding of spins in pairs halves the number of spin wave modes (see Section \ref{sec:disc}. Discussion).

%The Ruddlesden-Popper series of compounds \cite{ruddlesden1957} have the general formula $A_{n+1}B_{n}C_{3n+1}$. Examples of this class may be found among the CMR manganites, with the form $A_{n+1}$Mn$_{n}$O$_{3n+1}$, where $A$ is a rare-earth or alkali earth metal. \pcmo\, is thus an $n=\infty$ example of this class. In all the analysis that follows, the six-fold twinning of PCMO was accounted for. The twinning is described by systematic permutation of the three pairs of principal axes in the pseudocubic perovskite unit cell.

In light of the discussion above of the origin of both the intra and inter dimer magnetic exchange along the zig-zags originating from double exchange in the DDEX description, a more realistic dimer model needs to allow for weak dimerization, that is, the case when the two exchange parameters are similar. In a recent time-of-flight (ToF) neutron spectroscopy study \cite{Johnstone-2012} of the bilayer manganite Pr(Ca$_{0.9}$Sr$_{0.1}$)$_{2}$Mn$_{2}$O$_{7}$ (PCSMO) the higher branches of the dispersion which are present in the Goodenough model but not in the ZP model were successfully measured. This enabled the authors to show that the Goodenough model clearly provides the better description of the ground state in this material. Furthermore, the observed dispersion required both second neighbor interactions J$_{F2}$ and J$_{F3}$ along the FM zig-zags, which was ascribed to indicating some itinerant electron character. The authors also considered an alternative weakly dimerized model, shown in Fig. \ref{fig:scheme_def}(c).  The model allows for alternating stronger and weaker nearest neighbor FM exchange along the zig-zags on a purely phenomenological basis. In PCSMO the differences between this dimer model and the Goodenough model could be resolved in favor of the latter by inspecting maps of the intensity $S(\mathbf{Q},\omega)$ at a particular key energy, namely the top of the lower band of excitations.

In this paper we present an extensive investigation of the spin excitation spectrum of \pcmo. This material was chosen since it is perhaps the canonical example of a CE-type half-doped manganite. Furthermore, the whole debate surrounding whether the Goodenough or ZP model is an appropriate description of the interactions in the ground state arose because of the pioneering work of Daoud-Aladine {\it et al} \cite{daoud-aladine2002} on PCMO close to half-doping. We compare our time-of-flight inelastic neutron scattering measurements qualitatively and quantitatively to each of the Goodenough, ZP and generalised weak dimer model shown in Fig. \ref{fig:scheme_def}(d). We consider in detail the dispersion relations and dynamical structure factor, $S(\mathbf{Q},\omega)$, elucidating the specific features in the data that allow us to distinguish between the models. Despite the complication of six-fold twinning we are able to show that the Goodenough model provides the best description of our data. It is our intention that this detailed description of how inelastic neutron scattering data can be analyzed to distinguish between the different scenarios that can arise in DDEX models will serve as a reference for future work on manganites away from half-doping, for which the calculations indicate a greater instability towards the formation of dimer-like ground states.

%In this paper we present an investigation of the spin excitation spectrum of \pcmo. This material was chosen since it is perhaps the canonical example of a CE-type half-doped manganite. Furthermore, the whole debate surrounding whether the Goodenough or ZP model is an appropriate description of the ground state arose because of the pioneering work of Daoud-Aladine {\it et al} \cite{daoud-aladine2002} on PCMO close to half-doping. We also wished to determine whether the conclusions of Johnstone {\it et al}  \cite{Johnstone-2012} were specific to the two-dimensional bilayer example of a half-doped manganite, or are in fact more general. We compare our ToF inelastic neutron scattering measurements qualitatively and quantitatively to each of the Goodenough, dimer and ZP models. We consider in detail the dispersion relations and dynamical structure factor, $S(\mathbf{Q},\omega)$, elucidating the specific features in the data that allow us to distinguish between the models. Despite the complication of six-fold twinning, we are able to show that the Goodenough model provides the best description of our data.

%======================================================================
%Experimental methods section

\section{Experimental Methods}{\label{sec:meth}}

The ToF neutron scattering measurements were performed on the ARCS chopper spectrometer at the Spallation Neutron Source, Oak Ridge National Laboratory \cite{Abernathy-2012}. Data were collected with incident neutron energies ($E_{i}$) of 35, 70, and 140\,meV, with the instrument's corresponding Fermi chopper frequencies chosen in order to give energy resolution $\approx0.05E_{i}$ (full-width half-maximum) at the elastic line. The sample used for these measurements comprised a single crystal of mass 1.57\,g, grown as described previously \cite{Tomioka-1996}. It was mounted in an aluminium can containing helium exchange gas. The can was then mounted on the cold finger of a closed-cycle refrigerator (CCR) and cooled to 5.5\,K, at which temperature all measurements were performed.  The sample was mounted with the $a$- and $c$-axes horizontal. For the data collected with $E_{i}=35$\,meV and $E_{i}=140$\,meV the sample was held in a single orientation, with the $c$-axis of the sample parallel to the incident neutron beam. The data in these configurations were collected for 4 hours and 29 hours respectively. We hereafter refer to measurements taken in this configuration as `single-shot' datasets.

\begin{figure}[h]
\includegraphics*[scale=0.40,angle=0]{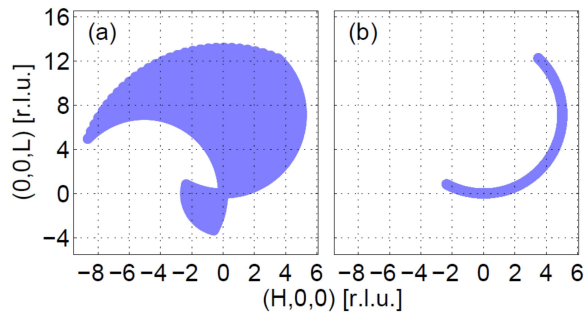}
\centering \caption{(Color online) (a) The reciprocal space coverage of the detectors in the (equatorial) $(H,0,0)/(0,0,L)$-plane, with $E_{i}=70$\,meV, zero energy transfer, and for a multi-angle dataset with the sample $c$-axis rotated stepwise from parallel to perpendicular to the incident beam. (b) Reciprocal space coverage for a single-shot dataset with the orientation fixed to be the $c$-axis parallel to the incident neutron beam.} \label{fig:angular_coverage}
\end{figure}

For the data collected with E$_{i}=70$\,meV, we performed a single-shot measurement with the $c$-axis of the sample parallel to the incident neutron beam, in which configuration we collected data for 14 hours. In addition we performed a measurement in which the sample $c$-axis was rotated about the vertical axis from parallel to perpendicular to the incident neutron beam, in 0.5$^{\circ}$ steps, with spectra recorded for $\sim1$\,hour at each orientation. The spectra at each orientation were then combined into a single large dataset using the {\sc Horace} software \cite{HoraceWeb}, allowing us to make a broad survey of the cross-section in the full 4-dimensional wavevector-energy space. We hereafter refer to measurements taken using this method as `multi-angle' datasets. Such a broad survey is crucial for a complete understanding of this material since, unlike bilayer PCSMO for example, the magnetic interactions have appreciable strength in all three dimensions. In a ToF measurement of this nature energy transfer is coupled to the component of the wavevector $\mathbf{Q}$ parallel to the incident neutron wavevector $\mathbf{k_{i}}$. Only by performing measurements in multiple orientations (or alternatively with multiple incident energies) can one decouple $\mathbf{Q}$ and energy \cite{HoraceWeb}.  To illustrate this, we show in Fig. \ref{fig:angular_coverage} the ARCS spectrometer's detector coverage of the $ac$-plane of reciprocal space for a single-shot run ($c$-axis parallel to the incident neutron beam) and for a $90^{\circ}$ angle scan with $E_{i}=70$\,meV.

\begin{figure}[h]
\includegraphics*[scale=0.4,angle=0]{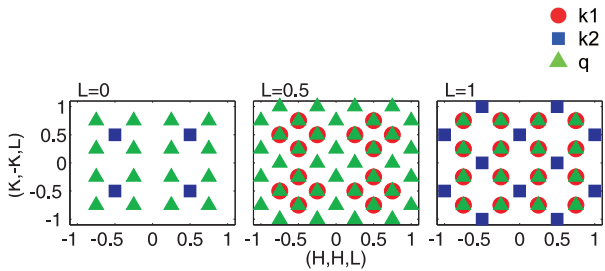}
\centering \caption{(Color online) The magnetic Bragg peaks $\mathbf{k_{1}}=(0,1/2,L)$ and $\mathbf{k_{2}}=(1/2,1/2,L)$, and structural superlattice peak $\mathbf{q} =(1/2,0,L^{\prime})$, indicated by red circles, blue squares and green triangles respectively. Here we have taken account of the six-fold structural twinning of the PCMO crystal. Notice that the $\mathbf{k_{2}}$ peaks do not overlap with the structural superlattice peaks, whereas the $\mathbf{k_{1}}$ peaks do. } \label{fig:twinfig}
\end{figure}

We have chosen to index reciprocal space according to space group $Pbnm$\cite{Lattice_note}. In this convention, the zig-zag chains propagate along the $(1,0,0)$-direction, as shown in Fig. \ref{fig:scheme_def}(a). This crystal structure is related to the cubic perovskite system, which has lattice parameter $a_{\rm{P}}=3.85\AA$, by $a\simeq b\simeq\sqrt{2}a_{\rm{P}}$, $c\simeq2a_{\rm{P}}$, thus the magnetic unit cell has size $2\sqrt{2}a_{\rm{P}} \times 2\sqrt{2}a_{\rm{P}} \times 2a_{\rm{P}}$. We therefore expect magnetic Bragg peaks, in the $Pbnm$ convention that we use, with wavevectors $\mathbf{k}_{1}=(0,1/2,L)$ and $\mathbf{k}_{2}=(1/2,1/2,L)$, and structural peaks with wavevector $\mathbf{q}=(1/2,0,L^{\prime})$, where $L$ is an odd integer and $L^{\prime}$ is any integer, arising as a result of a Jahn-Teller distortion of the Mn-O octahedra and orbital ordering \cite{radaelli1997}. In PCMO we expect sixfold twinning of the crystal that is described by a systematic permutation of the principal axes of the pseudocubic lattice. In order to demonstrate the effect of the twinning we show in Fig. \ref{fig:twinfig} the various peaks we expect to observe for a series of planes with $L=0$, 0.5 and 1 respectively. In the analysis of our INS data we will account for the twinning in the modeling explained in Section \ref{sec:disc}. In addition, we expect to observe crystal electric field transitions originating from the Pr ions. Inelastic neutron scattering experiments \cite{Krishnamurthy-2006} on Pr$_{0.5}$Sr$_{0.5}$MnO$_{3}$ show that there are transitions with energies 12.8\,meV and 15.35\,meV. We did not observe strong transitions (compared to the other sources of magnetic scattering) at these energies, however.

%=============================================
%Results section

\section{Results}{\label{sec:res}}

\begin{figure}[h]
\includegraphics*[scale=0.53,angle=0]{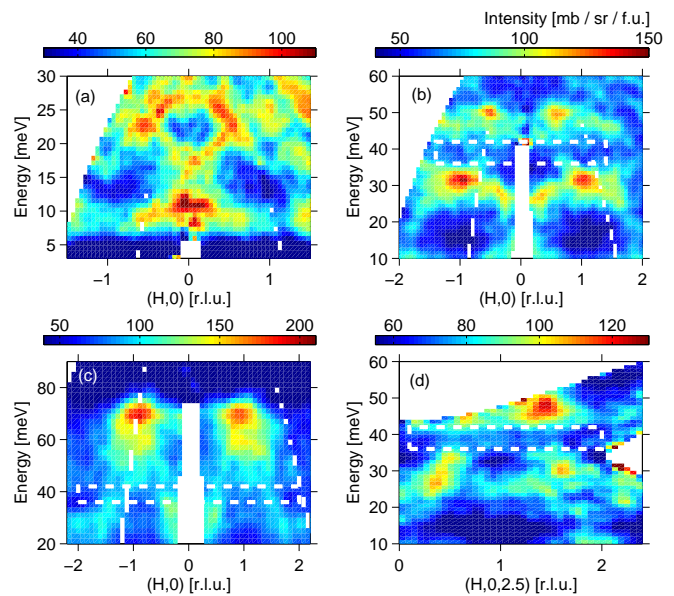}
\centering \caption{(Color online). Neutron scattering intensity maps of the dispersion along the $(H,0)$-direction, which is parallel to the zig-zag chains. The maps are taken from the single-shot datasets with (a) $E_{i}=35$\,meV, (b) $E_{i}=70$\,meV and (c) $E_{i}=140$\,meV.  The data have been smoothed by convolution with a Gaussian with FWHM equal to two bin sizes along each dimension for clarity. The map in panel (d) is taken from the multi-angle dataset with $E_{i}=70$\,meV, hence why the value of $L$ can be explicitly given. The white dashed rectangles in panels (b)-(d) indicate the position of the gap between upper and lower branches of the excitations, which is consistent with the Goodenough model. Note that the data in all four panels have been multiplied by $f(E)=\frac{E/E_{0}}{1-e^{-E/E_{0}}}$, where $E_{0}=k_{\rm{B}} \times 10$\,K. $f(E)$ is therefore constant when $E\ll E_{0}$ and $f(E)\propto E$ when $E\gg E_{0}$. This rescaling is to display the full spectrum more clearly on a single intensity scale. For panel (a) the signal was integrated over $\pm 0.05$ r.l.u. in the $(0,K)$ direction, whereas for panels (b)-(d) the integration was $\pm0.1$ r.l.u.} \label{fig:QE_overview}
\end{figure}

Figure \ref{fig:QE_overview} shows an overview of the dispersion from measurements with all three incident energies used. The essential features shown here are that the dispersion is highly structured over the full energy range, but most importantly there are two bands of dispersive excitations with an energy gap between them. Qualitatively, we estimate the two bands to cover energies in the range $0\lesssim E \lesssim 37$\,meV and $42 \lesssim E \lesssim 75$\,meV.  The data shown in panels (a) - (c) are taken from single-shot datasets, so the component of $\mathbf{Q}$ parallel to the incident neutron beam, $(0,0,L)$, varies as a function of energy. This makes qualitative statements difficult to make, other than the obvious presence of two bands of spin waves, but the value of $L$ is known for each energy so these data can be used collectively to constrain the parameters of different spin wave models. In panel (d) we show data from the multi-angle dataset, so the value of $L$ is the same for all of the data presented. This serves to illustrate that the gap observed between the lower and upper bands of the spin wave dispersion is not an artefact of the way in which a gapless dispersion might project on to the curved $(\mathbf{Q},E)$ hypersurface of measured reciprocal space coordinates from a single-shot run.

\begin{figure}[h]
\includegraphics*[scale=0.89,angle=0]{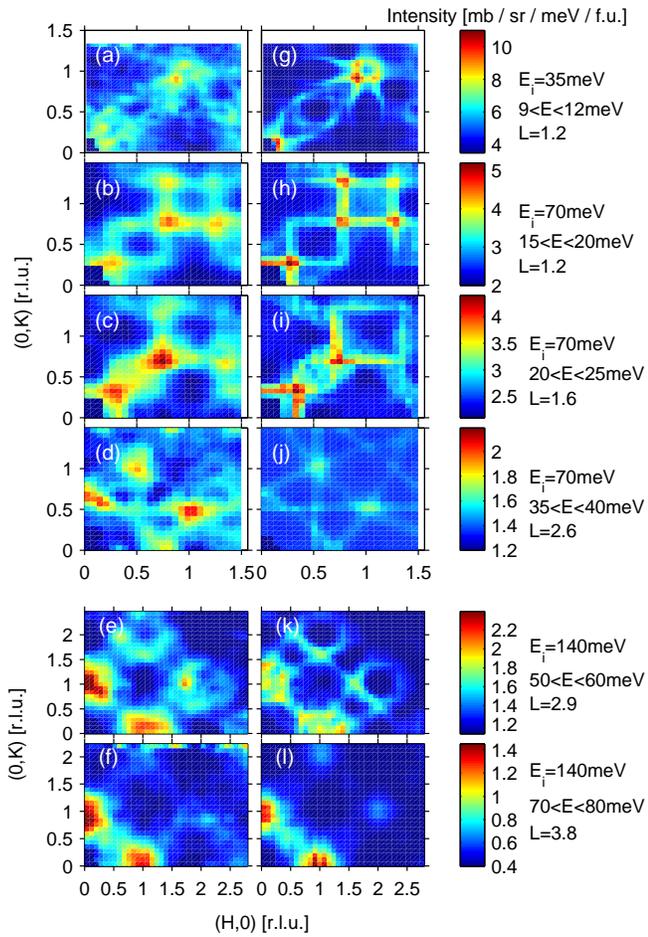}
\centering \caption{(Color online).  Panels (a) -- (f) show intensity maps taken with energies in the range 9$\leq${\it E}$\leq$12, 15$\leq${\it E}$\leq$20, 20$\leq${\it E}$\leq$25, 35$\leq${\it E}$\leq$40, 50$\leq${\it E}$\leq$60, and 70$\leq${\it E}$\leq$80\,meV, and incident energies of 35, 70, 70, 70, 140, and 140\,meV respectively. The data have been smoothed by convolution with a Gaussian with FWHM equal to two bin sizes along each dimension for clarity. Panels (g) -- (l) show corresponding simulations of those data using the Goodenough model and the global best fit parameters (see main text). All of the data shown here are from single-shot datasets with the $c$-axis parallel to the incident beam. We have made use of the symmetry of the crystal, and the symmetry and extent of the detector array by folding the data about the $(1,0,0)$ and $(0,1,0)$ axes to effectively increase the overall count rate by a factor of four. The calculated $S(\mathbf{Q},\omega)$ were convoluted with Gaussians with FWHM matched to the spectrometer's energy resolution at each energy and $E_{i}$ .} \label{fig:ConstE_overviewCE}
\end{figure}

To follow the evolution of the spin waves with energy, we show in Fig. \ref{fig:ConstE_overviewCE} a series of intensity maps at fixed energy from single-shot runs, together with simulations using the Goodenough model with the global best fit parameters obtained in the next Section. Because the value of $L$ changes with incident energy and energy transfer for a given $\mathbf{Q}$ coordinate in these plots, we are mapping the $L$ dispersion as well as the in-plane dispersion. The first four panels ((a) to (d)) therefore show a spin wave which disperses from $(1,1)$, with the fourth panel, (d), showing the top of the lower band of the spin waves. Panels (e) and (f)  show the upper band of spin waves, with the dispersion converging on $(1,0)$ as the band maximum at $\sim75$\,meV is approached. This part of the spectrum can be understood, with reference in addition to the periodicity evident in Fig. \ref{fig:QE_overview}(c), as arising from dominant FM exchange between nearest neighbor Mn ions along the zig-zags.

We incidentally note that the gross features described above are not dependent on a consideration of the structural twinning that is present in this material. That is to say, the presence of a gap must be a feature of all the twins present in our sample, and is independent of their relative populations. Similarly, the periodicity of the observed signal in momentum space can only arise from the periodicity of the spin chains. This is as opposed to a situation where contributions from different twins that might each have different periodicity that taken together combine to give an overall signal with periodicity not seen in any single twin.

\section{Discussion}{\label{sec:disc}}

The DDEX model \cite{vdBrink-1999,efremov2004}, which is a more general version of double exchange, would be an appropriate starting point for analyzing our data. However, calculation of the excitation spectra within such a framework is, though possible, far from straightforward \cite{Shannon-2002,Lv-2010}. In the absence of such calculations, we therefore use an effective Heisenberg Hamiltonian to model our data, just as was done for the half-doped manganites mentioned in the introduction \cite{Ulbrich2011,senff1,Johnstone-2012}.The use of an effective Heisenberg Hamiltonian is a well-established procedure to analyze the spin wave spectra in magnetically ordered systems, and not just those with purely integer or half-integer spins, nor even solely insulators. The exchange interactions in the model Hamiltonian that are required to reproduce the observed dispersion relations and intensities establish the strength, range and symmetries of the magnetic interactions. Examples where such an analysis is done include the parents of iron based superconductors \cite{Ca122,FeTe} in which the signs and magnitudes of the nearest neighbor and further neighbor exchange constants elucidated the dominant second neighbor antiferromagnetic exchange. In the case of manganites examples include the ferromagnetic metallic-like materials with $x\sim0.2-0.48$ (e.g. refs. \onlinecite{Hwang-98,Ye-2006,Perring-2001,hirota1}), where softening of the dispersion towards the zone boundary and the doping dependence of intra-bilayer exchange could be qualitatively explained by mapping the excitations of the double exchange model onto an equivalent ferromagnetic Heisenberg Hamiltonian (valid at the quasi-classical level) with corrections beyond $1/S$ and incorporating the $e_{g}$ orbital occupancies \cite{Shannon-2002,Khaliullin-2000,Jackeli-2002}.

To analyze the data quantitatively we calculated the dispersion relations and dynamical structure factors for the Heisenberg Hamiltonian $\mathcal{H}=- \sum_{<ij>}\rm{J}_{ij} \mathbf{S}_{i}\cdot\mathbf{S}_{j} - D(S_{i}^{z})^{2}$ using linear spin wave theory for each of the Goodenough, ZP and dimer models illustrated in Fig. \ref{fig:scheme_def}(a-d). In the present analysis appropriate choice of exchange interactions yields the known magnetic ground state for all of the models. The procedure for calculating the dispersion relations for the first two cases is explained in refs. \onlinecite{sikora2004,Ventura-2003}, in which textbook linear spin wave theory \cite{Walker-book} is applied to this particularly complex example. An extended explanation, including calculation of the structure factors, and will be the subject of a future publication. For the generalized dimer model the linear spin wave theory dispersion relations and dynamical structure factors were calculated using the McPhase package \cite{Mcphase-ref}. We confirmed that in the special case when this model reduces to the Goodenough model that it reproduces the results of ref. \onlinecite{sikora2004}. In the first part of the analysis procedure for each model, we obtained robust initial estimates for exchange constants from examination of the periodicities and energies of key spectral features detailed below. We then fitted the calculated $S(\mathbf{Q},\omega)$, averaged with equal weight from each twin, to the data by least-squares refining the exchange constants J$_{ij}$. The term $D$ is a single-ion anisotropy term to constrain the spins to the $ab$-plane. This term was included with a fixed value of 0.07\,meV in all of the models discussed in our analysis, to be consistent with the literature \cite{Johnstone-2012,Ulbrich2011,senff1}. Its small value means that it has negligible effect on the cross-section at the energies that we probed in our experiment and hence does not affect any of the results of our fits or our conclusions. The data that we fitted were all either one-dimensional cuts or two-dimensional slices taken at constant energy, so that the energy-dependent resolution of the spectrometer could be accounted for by using a fixed Gaussian broadening of the calculated intensity, the width of which was chosen to match the resolution at each energy. Twenty eight one-dimensional cuts and twelve two-dimensional slices were used in the fitting process, with the fit being performed simultaneously on all of them to determine a single set of best-fit parameters. The robustness of the fit parameters obtained was checked by using numerous different sets of starting values, to confirm that the fit always converged on the same result.

At this point we reiterate the distinction between the different models we will consider. In all cases they have the same magnetic ground state, as shown in Fig. \ref{fig:scheme_def}, which is commonly referred to as the CE magnetic structure. What distinguishes what we refer to as the `Goodenough model' and the `ZP model' and dimer models is thus the nature of the electronic environments around the Mn ions, which give rise to different sets of exchange interactions between them. Our INS measurements, and analysis with the effective Heisenberg model, therefore provide insight into the nature of this aspect of the ground state, rather than the magnetic structure itself, which is well established \cite{jirak}.

\subsection{Goodenough model}{\label{subsec:CE}}

We start with the Goodenough model. We use this term to apply to a model which has the CE magnetic structure and a set of exchange parameters as shown in Fig. \ref{fig:scheme_def}(a). Although we consider during our analysis the possibility of charge disproportionation, we do not {\it a priori} assume that it is present. The orbital order is implicity assumed to be that shown in Fig. \ref{fig:scheme_def}(a), since this then provides a natural explanation of how the signs of the nearest neighbor exchange interactions we consider come about. We also allow for the possibility of next-nearest-neighbor interactions along the FM zig-zags, as has been required in previous studies of similar systems \cite{Johnstone-2012,senff1,Ulbrich2011} and is taken to indicate a degree of electronic itinerancy along the zig-zag chains.

To simplify the analysis we initially assumed that the charge disproportionation between notional Mn$^{3+}$ and Mn$^{4+}$ sites is zero. This meant that all spins were set to an average value of $S=7/4$. It is well known that the charge disproportionation is substantially smaller than the ideal value \cite{goff2004,radaelli1997,jirak} and close to Mn$^{3.4+}$ and Mn$^{3.6+}$. By choosing to use equal spins on all sites, we are able to draw conclusions more straightforwardly about the magnitude and sign of exchange interactions from qualitative analysis of the periodicity and extent of the dispersion. Later in our analysis we will revisit this assumption and will show that the effect of introducing a magnetic moment disproportionation equal to the experimentally determined value \cite{jirak} does not alter the principal conclusions.

As has already been noted, the dominant exchange is the nearest-neighbor interaction J$_{F1}$ linking adjacent Mn ions along the zig-zag chains. Examination of Fig. \ref{fig:QE_overview}(c) shows that the maximum of the dispersion, which has the periodicity of a ferromagnet, is $\sim75$\,meV, and thus the strongest constraint on the size of J$_{F1}$ comes from this part of the dispersion. We determined from our global best fit that J$_{F1}=9.01\pm0.02$\,meV. This value is comparable to that measured in PCSMO ($11.39\pm 0.05$\,meV)\cite{Johnstone-2012}, La$_{0.5}$Sr$_{1.5}$MnO$_{4}$ (9.98\,meV)\cite{senff1}, the AFM CO phases of Nd$_{0.5}$Sr$_{0.5}$MnO$_{3}$ (6.76\,meV)\cite{Ulbrich2011}, and a little larger than the typical values for the FM double exchange in metallic FM manganites \cite{hirota1,Zhang-2009}. In other words, it is consistent with its origin being due to double exchange.

\begin{figure}[h]
\includegraphics*[scale=0.49,angle=0]{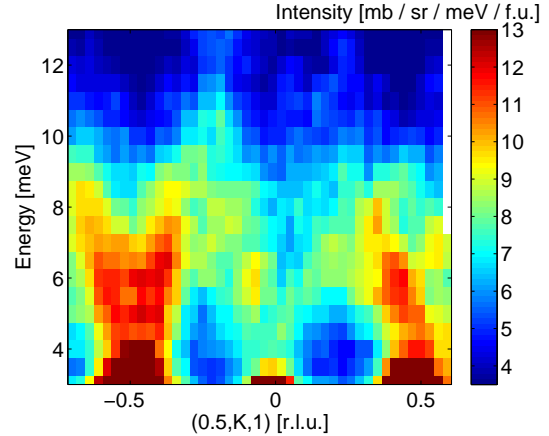}
\centering \caption{(Color online). Low energy dispersion, the extent in energy of which allows a reasonable estimate of J$_{A}$ to be obtained before the global fitting procedure.} \label{fig:QEslice_lowE_JA}
\end{figure}

At much lower energies ($\sim 10$\,meV) there is a branch of the dispersion, shown in Fig. \ref{fig:QEslice_lowE_JA}, with double this periodicity which arises from the antiferromagnetic coupling between zig-zag chains, J$_{A}$. Variations in this parameter have negligibly small effects on features of the dispersion above $\sim 10$\,meV, so in practice it can be determined from fits to the $E_{i}=35$\,meV dataset alone, where the background arising from the tails of the elastic line at these low energies is minimized. The global best fit yielded a value of J$_{A}=-0.45\pm0.01$\,meV.

\begin{figure}[h]
\includegraphics*[scale=0.49,angle=0]{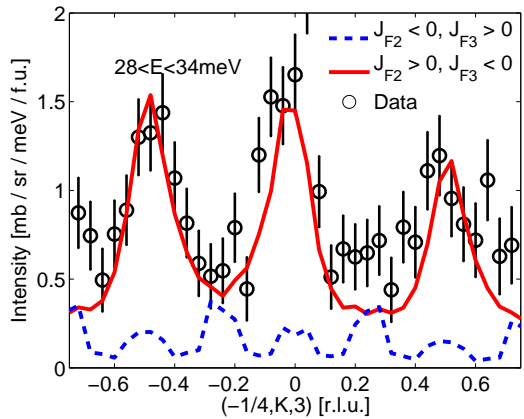}
\centering \caption{(Color online). Cut across part of the top of the lower band of excitations. The red (solid) line shows the global best fit to the data in this region, whereas the blue (dashed) line shows the structure factor if the signs of both J$_{F2}$ and J$_{F3}$ are reversed. Such a reversal preserves the magnitude of the gap between the lower and upper bands of excitations, and also the dispersion relation, but changes the structure factor as shown.} \label{fig:Qcut_J2J3_switch}
\end{figure}

The existence of a gap in the dispersion at $(1/2,0)$ (Fig. \ref{fig:QE_overview}(d)) immediately implies a halving of the periodicity in real space along the zig-zags. In the framework of an effective Heisenberg Hamiltonian this implies one or both of the second neighbor interactions J$_{F2}$ and J$_{F3}$ is non-zero. The position in energy and size of the gap between the lower and upper spin wave bands, shown schematically in Fig. \ref{fig:olga_dispersion}, define two energies which determine the values of both the next-nearest-neighbor terms J$_{F2}$ and J$_{F3}$. In order to reproduce the measured values J$_{F2}$ and J$_{F3}$ must both be non-zero, and we find that in fact they must have opposite signs, as was similarly concluded for PCSMO\cite{Johnstone-2012}. The size and position of the gap alone can be reproduced identically by exchanging the signs of J$_{F2}$ and J$_{F3}$, that is, with J$_{F2}<0<$J$_{F3}$ or J$_{F2}>0>$J$_{F3}$. However the details of the structure factor around the gap can distinguish between these two scenarios, specifically spectral weight is swapped between two relatively flat branches of the dispersion that lie at the top of the lower set of spin wave bands and the bottom of the upper set of spin wave branches. Also, the periodicity of peaks in the structure factor at the top of the lower set of spin wave branches is doubled along a key symmetry direction for the case of J$_{F3}>0>$J$_{F2}$, at variance with the data. This is illustrated in Fig. \ref{fig:Qcut_J2J3_switch}. For the global best fit we find that J$_{F2}=0.93\pm0.02$\,meV and J$_{F3}=-0.97\pm0.02$\,meV. Note that J$_{F3}$ is therefore antiferromagnetic.

A particular difference between PCMO and PCSMO is that in the former the coupling between planes is expected to be significant, whereas in the latter it is rather small. Indeed, previous low energy INS measurements of Nd$_{0.5}$Sr$_{0.5}$MnO$_{3}$ (ref. \onlinecite{Ulbrich2011}), which one would expect to be similar to PCMO, found that the inter-plane coupling was about a factor of four larger than the coupling between zig-zag chains. The complex twinning, together with an intermediate strength coupling between planes, means there is not a single feature which can be used to determine J$_{c}$ easily. The strength and position of the signal around $(\pm0.5,0)$ at $\sim 22$\,meV in Fig. \ref{fig:QE_overview}(a), with $E_{i}=35$\,meV, exhibits qualitative dependence on J$_{c}$, but overall the best method for refining J$_{c}$ is to use all of the data when performing the final global fit. We found J$_{c}=-2.09\pm0.03$\,meV, which like in Nd$_{0.5}$Sr$_{0.5}$MnO$_{3}$ is about a factor four larger than the in-plane antiferrogmagnetic coupling between zig-zag chains.

Let us now return to the question of charge / spin disproportionation between the different Mn sites. The effect of this being non-zero would be to have different values of the spin on alternate Mn sites along the zig-zags, which would halve the real-space periodicity and thereby open up a gap to create lower and upper dispersion branches, even with J$_{F2}=$J$_{F3}=0$.

To account for different moments on the Mn sites we can set $S=1.75+\delta S$ on the notional Mn$^{3+}$ sites and $S=1.75-\delta S$ on the notional Mn$^{4+}$ sites, instead of the uniform values of $S=1.75$ used up to now. In fact, the dispersion relations throughout the Brillouin zone for our global best fit parameters for the Goodenough model can be almost identically reproduced with J$_{F2}=$J$_{F3}=0$, but this requires $\delta S=0.42$. This corresponds to a ratio of moments on the Mn$^{3+}$ / Mn$^{4+}$ sites of 1.63, significantly at odds with the measured ratio \cite{jirak} of $3.18 \mu_{B} / 2.75\mu_{B} =1.16$. Alternatively, if the difference in the spins is considered to arise from charge disproportionation in a purely ionic model (in which Mn$^{3+}$ has $S=2$ and Mn$^{4+}$ has $S=3/2$) this corresponds to Mn$^{2.7+}$ and Mn$^{4.3+}$. In contrast, bond valence sums for PCMO from joint refinement of neutron and x-ray diffraction data \cite{goff2004} yield Mn$^{3.46+}$ and Mn$^{3.51+}$, consistent with typical disproportionation of $\lesssim 0.1e$ in half-doped manganites \cite{radaelli1997,goff2004,Herrero-Martin2004,garcia2001}. The gap therefore cannot be accounted for by the experimental moment or charge disproportionation. The contribution that it can make can be determined as follows. If we fix the ratio of the spins to that of the measured magnetic moments, then this changes the values of J$_{F2}$ and J$_{F3}$ to $0.70$\,meV and $-0.64$\,meV c.f. $0.93$\,meV and $-0.97$\,meV respectively, a reduction in the values of about 30\%. Alternatively, fixing the ratio of the spins by the values corresponding to typical charge disproportionation of $\pm 0.1e$ changes J$_{F2}$ and J$_{F3}$ to $0.84$\,meV and $-0.84$\,meV, a reduction compared to the values  assuming no disproportionation of less than 14\%. To summarize, the difference between the Mn moments or charge disproportionation on the two sites can therefore only account for a small fraction of the measured gap. The position in energy and the size of the gap therefore robustly requires non-zero J$_{F2}$ and J$_{F3}$ in the Goodenough model. We additionally note that accounting for the small charge / spin disproportionation as discussed above only results in a change of the values of J$_{F2}$ and J$_{F3}$ ; the other fitted exchange parameters, J$_{F1}$, J$_{A}$ and J$_{c}$, remain unaltered within the size of the errorbars.

\begin{figure}[h]
\includegraphics*[scale=0.49,angle=0]{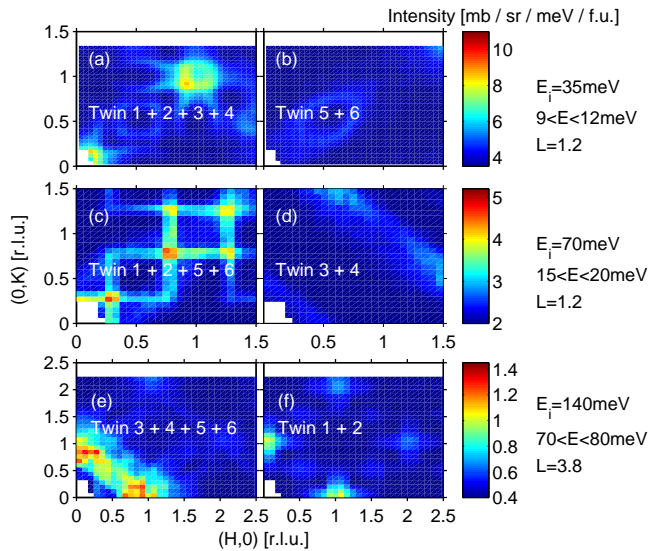}
\centering \caption{(Color online). Decomposition of some of the fits presented in Fig. \ref{fig:ConstE_overviewCE} into contributions from specific sets of twins. (a) Contribution from twins 1-4, which should be compared to Fig. \ref{fig:ConstE_overviewCE}(a) and (g); (b) contribution from twins 5 and 6, i.e. those which are not included in panel (a); (c) contribution from twins 1, 2, 5, and 6, which should be compared to Fig. \ref{fig:ConstE_overviewCE}(b) and (h); (d) contribution from twins 3 and 4, i.e. those not included in panel (c); (e) contribution from twins 3-6, which should be compared to Fig. \ref{fig:ConstE_overviewCE}(f) and (l); (f) contribution from twins 1 and 2, i.e. those not included in panel (e). Our convention for labeling the twins is described in the main text.} \label{fig:ConstE_twinfig}
\end{figure}

Our analysis has up to this point has also assumed an equal population of the six structural twins that are possible in this system. It is simplest to describe the twins with respect to the pseudo-cubic perovskite unit cell. Twin 1 is the unrotated coordinate system, and twin 2 corresponds to a $90^{\circ}$ rotation of this system about its pseudo-cubic $c$-axis. Twin 3 is a $90^{\circ}$ rotation of twin 1 about its $a$-axis, and then twin 4 is equivalent to twin 3 rotated by $90^{\circ}$ about the $c$-axis of twin 1. Twin 5 is a $90^{\circ}$ rotation of twin 1 about its $b$-axis, and finally twin 6 is a further $90^{\circ}$ rotation of twin 5 about the $c$-axis of twin 1.

We now illustrate the validity of the approach of assuming an equal population of twins in two ways. First, in Fig. \ref{fig:ConstE_twinfig} we show simulations of some of the constant-energy slices that have already been shown in Fig. \ref{fig:ConstE_overviewCE}. Specifically, panels (a) and (b), and (c) and (d), should be respectively compared to panels (g) and (h) of Fig. \ref{fig:ConstE_overviewCE}, and panels (e) and (f) should be compared to panel (l) of Fig. \ref{fig:ConstE_overviewCE}. We have re-calculated the scattering intensity for the global best fit to the Goodenough model, decomposed into the contribution from each of the six twins, and then formed combinations of them to show that there is no permutation of fewer than all six twins that will allow us to reproduce the fits to our dataset across the entire energy range. For example, in Fig. \ref{fig:ConstE_twinfig}(a) we can qualitatively reproduce the global best fit with contributions from the twins we label $1-4$, whereas in Fig. \ref{fig:ConstE_twinfig}(c) we require contributions from twins 1, 2, 5 and 6, and in Fig. \ref{fig:ConstE_twinfig}(e) all of the features in the data can be accounted for with just twins $3-6$. Second, we repeated the entire fitting procedure with additional fit parameters of the relative contributions from each twin. We found that, irrespective of the initial values of the fit parameters chosen, within the error on the fit parameters the populations of all six twin were essentially equal.

\subsection{Zener polaron model}{\label{subsec:ZP}}

Let us now turn to the ZP model (that is, of strongly bound dimers) for the sake of completeness, before considering the more realistic dimer model in Sec. \ref{subsec:dimer}. As shown in Fig. \ref{fig:olga_dispersion}, a key feature that distinguishes the Goodenough and ZP models is the presence or absence of a second band of excitations. By far the dominant exchange constant in the ZP model is expected to be that within the dimers. Accordingly, we expect one set of spin wave branches associated with coupling of the rigid ZP dimers that will have a band width determined by the characteristic inter-dimer exchange energies. These will be separated from a second set of branches associated with excitation of the pair of spins within the ZP by an energy gap determined by the much stronger intra-dimer exchange. That is, we expect a gap that could be several multiples of the bandwidth of the lower energy set of spin wave branches, which in the limiting case of the ratio of intra-dimer to inter-dimer exchange tending to infinity eliminates half the spin wave branches. In contrast, the Goodenough model permits a small (or even zero) gap between two sets of branches depending on the existence (or otherwise) of second neighbor interactions along the ferromagnetic zig-zag chains. As illustrated in Fig. \ref{fig:olga_dispersion}, judicious choice of J$_{A}$, J$_{FU}$ and J$_{FD}$ in this limiting ZP model (the model shown in Fig. \ref{fig:scheme_def}(b)) results in spin wave branches that closely follow the lower energy half of the spin wave branches of the Goodenough model, but in general the reverse is not possible. This is because the limiting ZP model has only half the number of spin wave modes by virtue of the rigid coupling of spins within the dimers. Moreover, the periodicity of the most energetic branch in our data corresponds to the nearest-neighbor Mn--Mn distance projected along the direction of the zig-zags. The nearest-neighbor separation of ZP dimers is twice that, which would result in a doubling of the periodicity of the spin wave branch compared to what we observe. In summary, because we have been able to measure the higher energy branches, which are separated from the lower branches by a relatively small gap, we are able to discount a model of weakly interacting dimers without making assumptions about the relative magnitudes of J$_{A}$, J$_{FU}$ and J$_{FD}$. In fact, this conclusion should apply to any case of interacting dimers, so long as the intra-dimer exchange is dominant over all other interactions (nearest neighbor and further neighbor) between a spin in one dimer and those in other dimers.

\begin{figure}[h]
\includegraphics*[scale=0.43,angle=0]{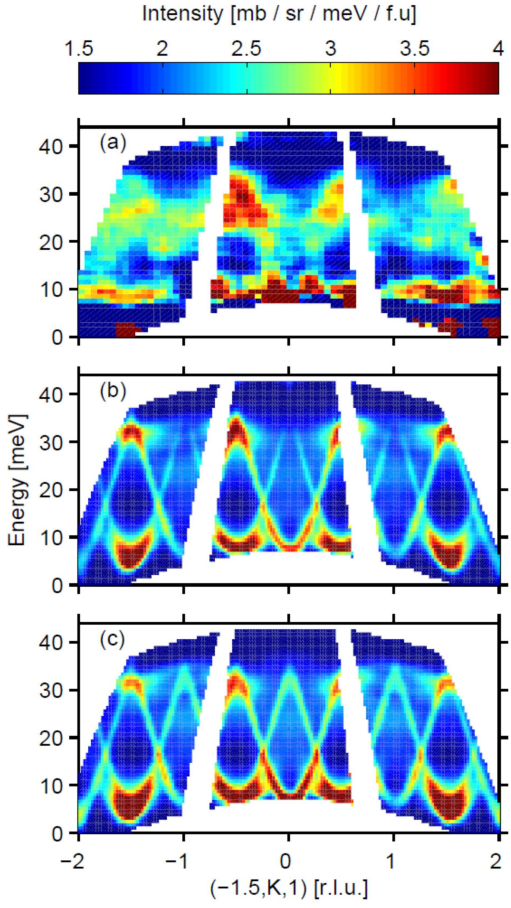}
\centering \caption{(Color online). (a) Intensity map of the dispersion along $(-1.5,K,1)$, taken from the $E_{i}=70$\,meV multi-angle dataset. The energy-dependent background arising from the tail of the incoherent elastic line has been subtracted. The signal was integrated over $\pm 0.1$ r.l.u. in the perpendicular $(H,0,0)$ and $(0,0,L)$ directions. (b) Simulation of the data in panel (a) using the global best fit parameters of the Goodenough model. (c) Simulation of the data in panel (a) using the global best fit parameters for the weak dimer model. These parameters give an essentially identical dispersion to the Goodenough model, with the only difference being the relative spectral weight at $(-1.5,0,1)$-type positions compared to that at $(-1.5,0.5,1)$-type positions at the top of the lower spin wave band.} \label{fig:QE_crucial}
\end{figure}

\subsection{Dimer model}{\label{subsec:dimer}}

As discussed earlier, it is more realistic to consider a model of weak dimerization as an alternative to the weakly interacting strongly bound dimers of the ZP model. A model along these lines was discussed by Johnstone {\it et al} in their work on PCSMO\cite{Johnstone-2012}, and is shown in Fig. \ref{fig:scheme_def}(c). The empirical rationale for this model was that the dominant exchange is the FM nearest-neighbor interaction along the zig-zags, but allowing for a small modulation between weaker and stronger exchange will create a gap at $(1/2,0)$ by virtue of doubling the repeat distance along the zig-zags. In PCSMO, with a suitable choice of parameters, this model produces an essentially indistinguishable dispersion relation to the Goodenough model, in addition to the accepted CE magnetic structure. However, the two models can be distinguished by differences in the structure factor at the top of the lower band of the dispersion.

The weak dimer model of ref. \onlinecite{Johnstone-2012} was purely phenomenological without reference to an electronic model. However, it is in fact a special case of a more general weak dimer model, shown in Fig. \ref{fig:scheme_def}(d), which we will use in the following discussion. The electronic justification for alternating stronger and weaker nearest-neighbor FM exchange J$_{FS}$ and J$_{FW1}$ arising from double exchange along the zig-zags has already been described in detail in the introduction. For the time being we disallow the next nearest neighbor exchange J$_{F2}$ and J$_{F3}$ along the zig-zags, but will include them later in this section. We also set all spins to an average value of $S=7/4$, as we did for the Goodenough model. In order to anticipate the likely signs of the weaker superexchange exchange interactions between zig-zags we need to consider the orbital character on the Mn sites. Model calculations for extensions to the DDEX Hamiltonian \cite{Barone2011,Yamauchi-JPSJ-2014} that include on-site Coulomb repulsion, MnO$_{6}$ tilting and Jahn-Teller distortions, consider the Mn orbitals to be a superposition of $d_{y^{2}-z^{2}}$ ($d_{x^{2}-z^{2}}$) and $d_{3x^{2}-r^{2}}$ ($d_{3y^{2}-r^{2}}$) character, where the $x$ and $y$ directions point along the two perpendicular Mn-Mn directions along the zig-zags. When the parameters of the model yield dimers there is a qualitatively closer resemblance to the latter on all sites. At the corner sites along the zig-zags the resemblance is less pronounced and tends to equal character (i.e. no orbital ordering) as the dimerization vanishes. Using the Goodenough-Kanamori rules \cite{goodenough1955,kanamori1959} for the expected superexchange between zig-zags, we might expect AFM coupling between nearest-neighbor Mn ions in parallel dimers, and FM coupling between nearest-neighbor Mn ions in perpendicular dimers. We therefore denote the AFM interaction between nearest-neighbors in parallel dimers with the exchange parameter  J$_{A}$, and the weak FM interaction between nearest-neighbors in perpendicular dimers with the exchange parameter J$_{FW2}$. This model differs from that proposed by Johnstone {\it et al}, in which it was assumed J$_{A2}\equiv$J$_{FW2}<0$. Note that although we use the Goodenough-Kanamori rules to anticipate the signs of J$_{A}$ and J$_{FW2}$, we do not actually constrain the signs in our modelling. Consequently, the dimer model shown in Fig. \ref{fig:scheme_def}(c) is a limiting case of the more general model shown in Fig. \ref{fig:scheme_def}(d).

We fitted the dimer model in the same manner as the Goodenough model. An estimate for the strong intra-dimer coupling J$_{FS}$ was made from the highest-energy part of the dispersion. The size of the gap between the upper and lower spin wave branches is mostly determined by the difference between the intra- and inter-dimer couplings along the zig-zags, which enabled an estimate for J$_{FW1}$ to be made. Estimates for the weaker couplings between the zig-zags, J$_{A}$ and J$_{FW2}$, were determined from low-energy features of the data. The dimer model parameters were then refined using the same cuts as were used for the refinement of the Goodenough model parameters. The global best fit parameters were then J$_{FS}=10.74\pm0.03$\,meV, J$_{FW1}=7.02\pm0.04$\,meV, J$_{FW2}=1.26\pm0.11$\,meV, J$_{A}=-2.47\pm0.14$\,meV, and J$_{c}=-2.19\pm0.03$\,meV. This produces an almost identical dispersion relation to that obtained from the global best fit parameters to the Goodenough model.

The small size of the gap between the spin wave branches compared to the overall bandwidth is reflected by rather weak dimerization, quantified by the difference between J$_{FS}$ and J$_{FW1}$ as a fraction of their average, being only 42\%. It is therefore not surprising that both J$_{FS}$ and J$_{FW1}$ are quite close in value to J$_{F1}$ in the Goodenough model, since they are largely determined by the maximum energy of the dispersion. As an interesting aside, the magnitude of the AFM coupling between the zig-zag chains in the best fit to the dimer model is a factor $\sim5.5$ larger than in the Goodenough model. This can be understood by noticing that J$_{A}$ is used for the two exchange pathways in the Goodenough model that are described by J$_{A}$ and J$_{FW2}$ in the dimer model. In order for the same magnetic structure to be stable in the dimer model, $|$J$_{A}|$ has to be larger than J$_{FW2}$, since there are two AFM contributions (J$_{A}$) to the total energy in the Goodenough model, balanced against one AFM contribution (J$_{A}$) and one FM contribution (J$_{FW2}$) in the dimer model. These last two terms effectively trade against one another to give an overall antiferromagnetic interaction between zig-zags and close to identical dispersion to that observed in the Goodenough model. The interaction between the planes that contain the zig-zag chains, J$_{c}$, is almost the same for the dimer and Goodenough models, since this term describes the same physics in both cases.

\begin{figure}[!h]
\includegraphics*[scale=0.54,angle=0]{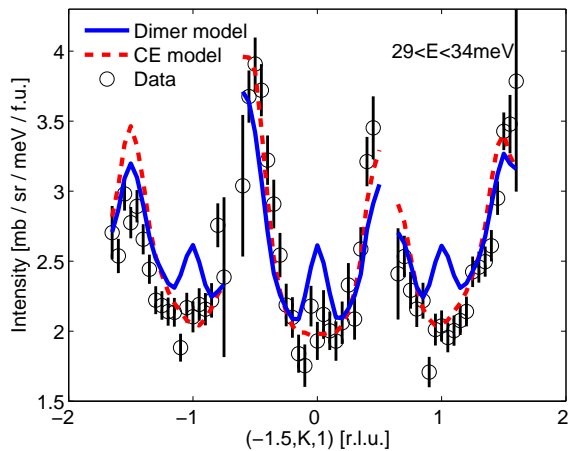}
\centering \caption{(Color online). Cut through the data shown in Fig. \ref{fig:QE_crucial}(a), integrating the signal over $29 < E < 34$\,meV. The data are indicated by black circles. The simulations using the Goodenough and dimer models are indicated by dashed (red) and solid (blue) lines respectively. } \label{fig:Qcut_crucial}
\end{figure}

One might be tempted to conclude that because the dispersions obtained from global best fits to both the Goodenough and dimer models are essentially indistinguishable, the models are equivalent. In fact this is not the case. For most of the energy scale of the excitations the two models have very similar structure factors. However we find that near the top of the lower spin wave band the structure factors are different, and are readily distinguished on examination of the data along a well-chosen symmetry direction. This difference in structure factor between the data and Goodenough model, and the dimer model, is illustrated in Fig. \ref{fig:QE_crucial}. The agreement between the data and Goodenough models (panels (a) and (b)) is good, but there is enhanced intensity around $(0.5,K,1)$ for integer $K$ arising from a larger structure factor from the dispersion which converges on these points in the dimer model (panel (c)) which does not agree with the data. To quantify this difference further we show in Fig. \ref{fig:Qcut_crucial} a one-dimensional cut across the top of the lower band of the dispersion. The dashed line is the Goodenough model global best fit, which clearly captures all of the essential features of the data, whereas the solid line is the dimer model global best fit, that fails to do so. The differences are clearly statistically significant, since the difference between the dimer model and the data at $(0.5,K,1)$ for integer $K$ is much larger than the size of the errorbars on the data points.

The essential physical difference between the dimer and Goodenough models is that the former allows for the gap at (1/2,0) and $\sim40$\,meV to be accounted for by allowing alternating nearest-neighbor FM exchange along the zig-zags, whereas the latter accounts for it using further neighbor terms along the zig-zags. A question which naturally arises is what is the result if both weak dimerization and further neighbor interactions are permitted. The absence of peaks in the structure factor at $(0.5,K,1)$ for integer $K$ shown in Fig. \ref{fig:Qcut_crucial} strongly supports the pure Goodenough model without any element of dimerization, as the peaks appear to be a feature solely of the dimerization. We can quantify this assertion by fitting the data in Fig. \ref{fig:Qcut_crucial} with a generalization of the dimer model that adds the same further neighbor terms that were used in the Goodenough model, J$_{F2}$ and J$_{F3}$. With the same exchange interactions between the FM zig-zags, the dispersion and structure factor of the Goodenough model are reproduced exactly in such a dimer model with the alternating nearest-neighbor FM exchange terms set to the same value i.e. zero dimerization, and the further neighbor terms equal to the Goodenough model values. For the fit, we fixed the overall bandwidth of the spin waves (which fixed the average of nearest neighbour FM exchange, (J$_{FS}+$J$_{FW1}$)/2)), the energies of the centre and size of the gap (which further constrains J$_{F2}=-$J$_{F3}$), and the inter-chain exchange terms, to the values from the global best fit dimer model, and then allowed the dimerization (the difference between J$_{FS}$ and J$_{FW1}$ as a fraction of their average) and J$_{F2}$($=-$J$_{F3}$) to vary. The result is J$_{F2}=-$J$_{F3}=0.84\pm0.09$\,meV, very close to the values for the global best fit to the Goodenough model in Sec. \ref{subsec:CE} (J$_{F2}=0.93\pm0.02$\,meV and J$_{F3}=-0.97\pm0.02$\,meV), and the dimerization is reduced from $42\%$ in the dimer model without the further neighbor terms to $6.5\pm6.6\%$. We therefore conclude that adding further neighbor exchange interactions to the dimer model permits the best fit to revert essentially to the Goodenough model.

%======================================================================

\section{Conclusion}{\label{sec:conc}}
We have studied the spin waves in PCMO using time-of-flight neutron spectroscopy and found that the upper and lower branches of the dispersion are gapped. The gapped spin wave dispersion is consistent with the Goodenough model with strong nearest-neighbor ferromagnetic interactions along the zig-zag chains and weak antiferromagnetic interactions between them, provided there exist small next-nearest-neighbor couplings J$_{F2}$ and J$_{F3}$ along the zig-zags. Gross features of the spin wave dispersion allow us to rule out the Zener polaron picture in the limit of strongly bound dimers as a description of the ground state of PCMO. An alternative model in which pairs of Mn ions are weakly dimerized, is shown to provide an equally good description of the spin wave dispersion relation as the Goodenough model, but there are significant differences between the spin wave intensities in this weak dimer model and the data. These differences do not arise in the Goodenough model, so our data indicate that the Goodenough model (an undimerized model) with significant second neighbor interactions within the FM zig-zags provides the most satisfactory description of the magnetic ground state of PCMO. The existence, sizes and signs of J$_{F2}$ and J$_{F3}$ are rigorously set by the gap in the spin wave dispersion. Any charge / moment disproportionation between the Mn sites along the zig-zags is only able to account for a small fraction of the gap.

We found that the nearest neighbor FM interaction along the zig-zags is comparable or larger than the typical values found for the FM exchange in metallic manganites \cite{hirota1,Zhang-2009}, a result consistent with previous studies of other half-doped manganites \cite{Johnstone-2012,senff1,Ulbrich2011}. The magnitudes of the second neighbor interactions along the zig-zags are about twice that of the antiferromagnetic exchange between zig-zags. The significant further neighbor interactions are suggestive of delocalization of the $e_{g}$ electrons along the zig-zags. These results are consistent with the theoretical analysis of the degenerate double exchange (DDEX) model for half-doping \cite{vdBrink-1999,efremov2004}. In this model, a cooperative ordering of the charge and degenerate $e_{g}$ orbital degrees of freedom takes place to accommodate the frustration between delocalization (which favors ferromagnetism) and the antiferromagnetic superexchange between the core $t_{2g}$ moments. The result is FM zig-zag chains that are antiferromagnetically coupled via superexchange, but with the FM exchange along the zig-zags arising from the double exchange mechanism. This explains the fact that the nearest neighbor FM exchange we measure in PCMO is comparable in magnitude to that found in metallic manganites. In the DDEX model the double exchange overcompensates for the antiferromagnetic superexchange along the zig-zags, and the delocalization of the $e_{g}$ electrons plausibly accounts for the second neighbor exchange along the zig-zags that exceeds the antiferromagnetic superexchange between the core $t_{2g}$ Mn moments. For $x=0.5$ the charge modulation of $\lesssim0.1e$ is site centered so that all Mn--Mn bonds along the zig-zag are identical. Extensions to the DDEX model that couple the electronic and lattice degrees of freedom allow for cooperative distortion or tilt of the MnO$_{6}$ units that can result in magnetic dimerization for a range of parameters in the model. However, our results indicate that any such coupling does not affect the magnetic exchange interactions in PCMO, and that the magnetic degrees of freedom are best described by the Goodenough model within the DDEX picture.

Given that all of the toy models we have considered provide a way to examine different ground states that come out of the DDEX model and extensions to it, our analysis can serve as a template for how to understand future experiments on other manganites away from half doping, where the predictions from DDEX suggest that the CE / Goodenough picture is less likely to apply. Given that the balance between the different ground states in these models is rather delicate, having a toolkit available which can distinguish between them in real materials with different doping will provide valuable input to further calculations.

We are grateful to A. Daoud-Aladine, G. A. Sawatzky, J. P. Hill and F. Kr\"{u}ger for helpful discussions, and to A. J. Wenban for work on the dimer spin wave model. This work was supported by the Science and Technology Facilities Council of Great Britain. O.S. acknowledges support by the Polish National Science Center (NCN) under Project No. 2012/04/A/ST3/00331. This research at Oak Ridge National Laboratory's Spallation Neutron Source was sponsored by the U.S. Department of Energy, Office of Basic Energy Sciences.

\bibliography{pcmo_bib2}

\end{document}